\documentclass[iop,usenatbib]{emulateapj-rtx4}
\usepackage{amsmath}
\usepackage{pifont}% http://ctan.org/pkg/pifont
\usepackage[usenames,dvipsnames]{xcolor}
\usepackage[normalem]{ulem}
\usepackage{cancel}
\usepackage{enumitem}

\newcommand{\sph}{{\sc sph}}   % SPH in small caps
\newcommand{\Msolar}{{\rm M_{\odot}}}   % solar mass symbol
\newcommand{\MJup}{M_{\rm Jup}}

\newcommand{\icarus}{Icarus}

\shorttitle{On the disc morphology in Elias 2--27}
\shortauthors{Meru et al.}

\begin{document}

%% LaTeX will automatically break titles if they run longer than
%% one line. However, you may use \\ to force a line break if
%% you desire.

\title{On the origin of the spiral morphology in the Elias 2--27 circumstellar disc}

%% Use \author, \affil, and the \and command to format
%% author and affiliation information.
%% Note that \email has replaced the old \authoremail command
%% from AASTeX v4.0. You can use \email to mark an email address
%% anywhere in the paper, not just in the front matter.
%% As in the title, use \\ to force line breaks.

%\author{ }
%\affil{ }
%\email{ }

\author{Farzana~Meru}
\email{farzana.meru@ast.cam.ac.uk}
\author{Attila~Juh{\'a}sz}
\email{juhasz@ast.cam.ac.uk}
\author{John~D.~Ilee}
\email{jdilee@ast.cam.ac.uk}
\author{Cathie~J.~Clarke}
\author{Giovanni~P.~Rosotti}
\author{Richard~A.~Booth}

\affil{Institute of Astronomy, University of Cambridge, Madingley Road, Cambridge, CB3 0HA, UK}

%% Notice that each of these authors has alternate affiliations, which
%% are identified by the \altaffilmark after each name.  Specify alternate
%% affiliation information with \altaffiltext, with one command per each
%% affiliation.

%\altaffiltext{1}{Visiting Astronomer, Cerro Tololo Inter-American Observatory.
%CTIO is operated by AURA, Inc.\ under contract to the National Science
%Foundation.}
%\altaffiltext{2}{Society of Fellows, Harvard University.}
%\altaffiltext{3}{present address: Center for Astrophysics,
%    60 Garden Street, Cambridge, MA 02138}
%\altaffiltext{4}{Visiting Programmer, Space Telescope Science Institute}
%\altaffiltext{5}{Patron, Alonso's Bar and Grill}

%% Mark off your abstract in the ``abstract'' environment. In the manuscript
%% style, abstract will output a Received/Accepted line after the
%% title and affiliation information. No date will appear since the author
%% does not have this information. The dates will be filled in by the
%% editorial office after submission.

\begin{abstract}

The  young star Elias  2-27 has  recently been  observed to posses a massive circumstellar  disc with two prominent large-scale spiral arms.  In this Letter we perform three-dimensional Smoothed Particle Hydrodynamics simulations, radiative transfer modelling, synthetic ALMA imaging and an unsharped masking technique to explore three possibilities  for  the  origin  of  the  observed structures -- an  undetected companion either internal or external to the spirals,  and  a self-gravitating  disc.  We find that a gravitationally unstable disc and a disc with an external companion can produce morphology that is consistent with the observations.  In addition, for the latter, we find that the companion could be a relatively massive planetary mass companion ($\lesssim 10 - 13\MJup$) and located at large radial distances (between $\approx 300 - 700$~au).  We therefore suggest that Elias 2-27 may be one of the first detections of a disc undergoing gravitational instabilities, or a disc that has recently undergone fragmentation to produce a massive companion.

\end{abstract}

%% Keywords should appear after the \end{abstract} command. The uncommented
%% example has been keyed in ApJ style. See the instructions to authors
%% for the journal to which you are submitting your paper to determine
%% what keyword punctuation is appropriate.

%\keywords{globular clusters: general --- globular clusters: individual(NGC 6397,
%NGC 6624, NGC 7078, Terzan 8}

%% From the front matter, we move on to the body of the paper.
%% In the first two sections, notice the use of the natbib \citep
%% and \citet commands to identify citations.  The citations are
%% tied to the reference list via symbolic KEYs. The KEY corresponds
%% to the KEY in the \bibitem in the reference list below. We have
%% chosen the first three characters of the first author's name plus
%% the last two numeral of the year of publication as our KEY for
%% each reference.

%% Authors who wish to have the most important objects in their paper
%% linked in the electronic edition to a data center may do so by tagging
%% their objects with \objectname{} or \object{}.  Each macro takes the
%% object name as its required argument. The optional, square-bracket 
%% argument should be used in cases where the data center identification
%% differs from what is to be printed in the paper.  The text appearing 
%% in curly braces is what will appear in print in the published paper. 
%% If the object name is recognized by the data centers, it will be linked
%% in the electronic edition to the object data available at the data centers  
%%
%% Note that for sources with brackets in their names, e.g. [WEG2004] 14h-090,
%% the brackets must be escaped with backslashes when used in the first
%% square-bracket argument, for instance, \object[\[WEG2004\] 14h-090]{90}).
%%  Otherwise, LaTeX will issue an error. 

\section{Introduction}

With the  advent of the Atacama Large Millimetre Array (ALMA), it
has for the first time become  possible to spatially resolve, and thus
directly observe, the midplane structure of protoplanetary discs where
planet   formation  processes   occur.   Such   an  extreme   increase  in
observational capability has given rise to several surprising results,
examples  of which  include the  symmetric ring  structures in  HL Tau
\citep{ALMA_HLTau}, TW  Hydrae \citep{Andrews_TWHydra} and HD 163296 \citep{isella_2016},  the
horseshoe shaped dust traps  in HD 142527 \citep{casassus_2013},
and the  birth of  a ternary  system via  disc fragmentation  in L1448
IRS3B  \citep{Tobin_Nature_GI_fragment}.  Observations  such as  these
are extremely  powerful, as sub-structure within  protoplanetary discs
can be a signpost of  dynamical or chemical effects occurring within the star-disc
system.  Therefore, a proper understanding of their cause is essential
to determine  which of these  processes are important during  the star
and planet formation process.

\smallskip

A  recent  example  of   such  a  spatially-resolved  observation  was
presented  by \citet{Elias227_Science} in  which \objectname{Elias  2-27} was targeted with ALMA.  Elias 2-27 is a  low mass young star  (M$_{\star} = 0.5$ -- $0.6$~$\Msolar$, t$_{\rm age}$~$\sim 1$~Myr;
\citealt{rho_oph_spectraltype_age,rho_oph_accretion}).  Based on its spectral energy distribution (SED), the system is thought to belong to the Class II phase \citep{andrews_2009_oph, evans_2009_c2d}, yet at the same time, observations have suggested an unusually large disc mass, ranging from 0.04 -- 0.14~$\Msolar$ \citep{andrews_2009_oph, isella_2009_structure, ricci_2010_oph}.  

\smallskip

\begin{figure*}
    \centering
    \includegraphics[width=\textwidth]{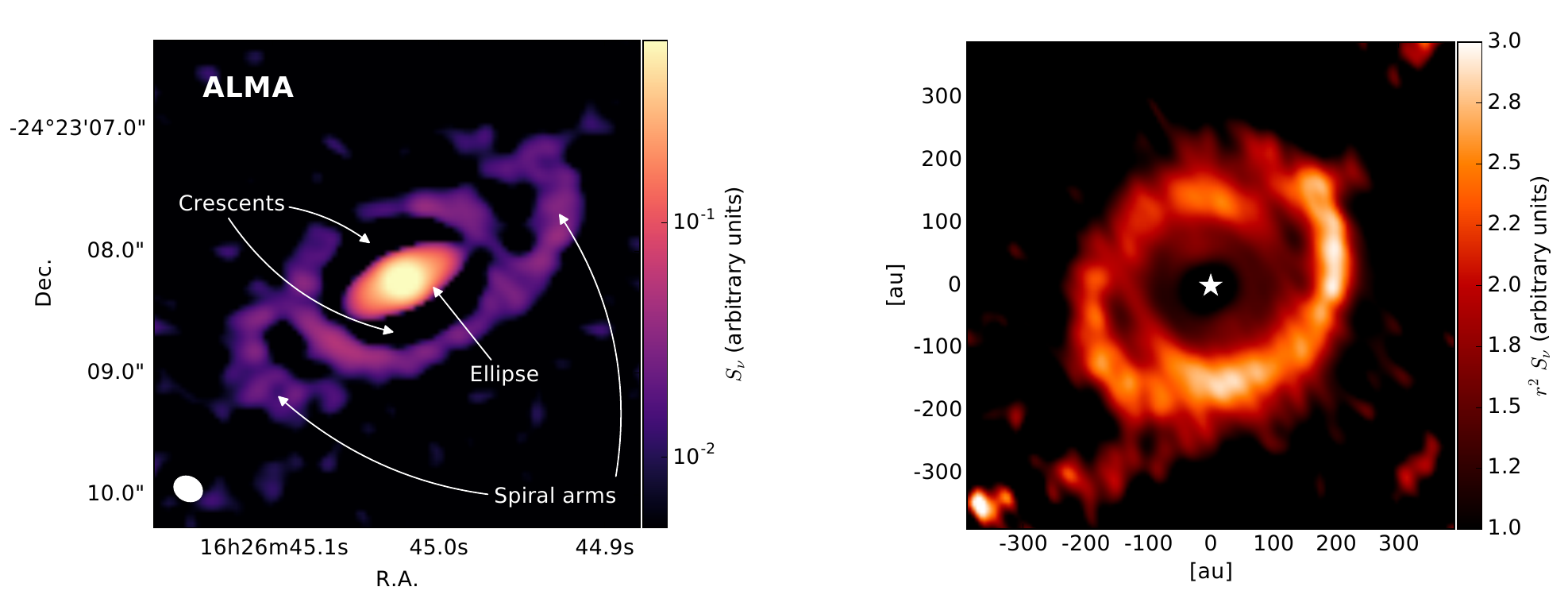}
    \caption{Left: 1.3\,mm continuum image of Elias 2-27, processed with an unsharp masking filter (originally presented by \citealt{Elias227_Science}).  Two symmetric spiral arms, a bright inner ellipse, and two dark crescents are clearly visible.  The beam is shown in the lower left corner as a filled ellipse.  Right: Deprojection of the original 1.3mm image with an $r^{2}$ scaling applied, showing two prominent spiral arms.  The white star denotes the central star's location.  The apparent ring structure and the central ellipse in the left image are artefacts of the unsharp masking.}
    \vspace{0.5em}
    \label{fig:ALMA1}
\end{figure*}

The Elias 2-27 disc posseses two large-scale symmetric spiral arms (Figure~\ref{fig:ALMA1}, right).  Additionally when the raw ALMA observations are processed with an unsharp masking filter, two dark crescents interior to the spirals and a bright inner ellipse are revealed (Figure~\ref{fig:ALMA1}, left).  The origin of these is unclear.  With such a large disc-to-star mass ratio, could the disc be self-gravitating?  Or, could an as-yet-undetected companion be causing these features via dynamical interactions?       

\smallskip

In this  Letter we describe  the results of hydrodynamical and radiative transfer modelling of the Elias  2-27 system.  We produce synthetic ALMA observations to explore three possibilities that may give rise to the observed features --- a  companion internal to  the spirals,  a  companion external to the spirals, or gravitational instabilities operating within the disc.

\section{METHODOLOGY}
\label{sec:numerics}

\subsection{Hydrodynamics}
\label{sec:hydro}

Our  simulations  are  performed  using a  three-dimensional  Smoothed Particle   Hydrodynamics  code   ({\sph}) which includes  the heating  due to  work done  and  the radiative transfer  of  energy in the flux-limited diffusion limit \citep{WH_Bate_Monaghan2005,WH_Bate_science}.   A detailed  description of the  code can  be found  in \cite{Triggered}, however  for  this  work  we  employ two  differences.   Firstly,  boundary particles are located at every timestep, allowing the vertical location between  the optically thick and thin region to be regularly re-evaluated,  leading to  more accurate  boundary temperatures.   Secondly,   we  employ  the  \cite{MorrisMonaghan1997} artificial  viscosity with  the {\sph}~parameter,  $\alpha_{\rm SPH}$, varying between 0.1 -- 1.0 and $\beta_{\rm SPH} = 2 \alpha_{\rm SPH}$, to model shocks within the disc.

\smallskip

We perform 72 hydrodynamical simulations varying a number of disc properties as well as orbit properties for the companion simulations.  We first describe our reference disc setup: we model a $0.5 \Msolar$ star surrounded by a disc whose temperature follows
\begin{equation}
T(R) = 13.4 {\rm K} \left ( \frac{R}{200~{\rm au}} \right )^{-q}.
\label{eq:T}
\end{equation}
The disc surface density follows
\begin{equation}
\Sigma(R) = \Sigma_{\rm c} \left ( \frac{R}{R_{\rm c}} \right )^{-p} {\rm exp} \left ( - \left ( \frac{R}{R_{\rm c}} \right )^{2-p} \right ),
\label{eq:sigma}
\end{equation}
where $\Sigma_{\rm c}$ is the surface mass density of gas at the cut-off radius, $R_{\rm c}$.  We model this disc between $R_{\rm in} = 10$~au and $R_{\rm out} = 400$~au using the parameters determined by \citet{Elias227_Science}, namely $\Sigma_{\rm c} = 5$\,g\,cm$^{-2}$, R$_{\rm c} = 200$\,au, $q = 0.45$ and $p=0.7$.

\smallskip

We then perform a suite of simulations with a wide range of parameters to test the internal companion, external companion and gravitational instability hypotheses.  Due to the computational expense of each simulation, our aim is not to fit the exact parameters, but to test whether each hypothesis can reproduce the observed morphology.  We vary the disc mass, surface mass density profile, temperature profile, cut-off radius, and outer disc radius.  We also model pure power surface density profiles, i.e. without the exponential term in equation~\ref{eq:sigma}, so that, once evolved, the disc has a much steeper profile in the outer regions.  In addition, for the companion simulations, we vary the companion mass, the pericentre distance, eccentricity and inclination.

\smallskip

Each disc is modelled using 250,000 {\sph} gas particles.  The  ratio of the smoothing length to the disc scale-height is $< 0.5$  outside
$40$ --  $60$~au,  giving sufficient resolution to probe dynamical effects on the disc.  The discs
are  modelled  with radiation  hydrodynamics using  flux-limited
diffusion, and the surface temperature (representing irradiation from the central star) is held at the profile given by Eqn.~\ref{eq:T}.  For the parameters studied here, the disc is optically thick to stellar irradiation, and remains vertically isothermal at the boundary temperature beyond a radius of 20 and 30~au in the companion and self-gravitating disc simulations, respectively.  This implies that the thermodynamics are mainly set by external irradiation.

We assume that the gas and dust are well mixed.  For our self-gravitating simulations the Stokes numbers of millimeter particles are $\mathcal{O}(0.01)$, and thus dust trapping in the spirals is expected to be marginal \citep{Booth_Clarke_GIdust,Shi_Chiang2014}.  In our simulations with a companion the Stokes numbers are $\mathcal{O}(0.1)$.  For a spiral to trap dust, however, it is also necessary that the timescale for it to concentrate towards the pressure maximum \citep{Clarke_GI_planetesimals} is less than the crossing timescale of a spiral feature.  Although this condition is readily met in the case of self-gravitating discs where the spirals nearly co-rotate with the Keplerian flow, this is not true in the case of a planet generated spiral which co-rotates with the planet \citep[e.g.][]{Paardekooper2006_dust,Zhu_dust_filtration,Birnstiel_dust_conc_time}

\begin{figure*}
\centering
\includegraphics[width=1.0\textwidth]{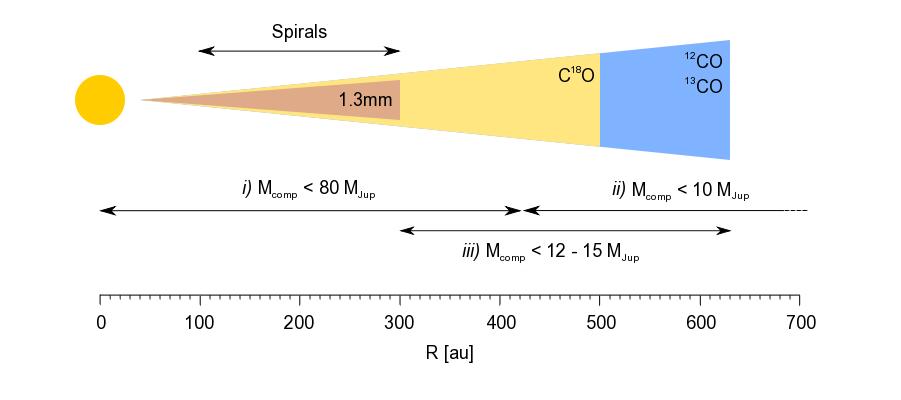}
\caption{Schematic diagram of the Elias 2-27 system based on the results of \citet{Elias227_Science}.  Also shown are the companion mass limits based on observations (i) by \cite{ratzka_2005}, (ii) from the UKIDSS data and (iii) theoretical gap-opening limits (see Section \ref{sec:mcomp}).}
\vspace{2em}
\label{fig:elias_diagram}
\end{figure*}

\subsection{Radiative Transfer \& Synthetic Imaging}
\label{sec:rt}

To  calculate synthetic observations of our models, we  use the  3D
radiative transfer code {\sc radmc-3d}\footnote{http://www.ita.uni-heidelberg.de/\~{}dullemond/software/radmc-3d/}.  The  dust  opacity  is  calculated  from  the  optical  constants  for
astronomical  silicates   \citep{weingartner_2001}  using  Mie-theory,
assuming  a power-law grain  size distribution  between 0.005  -- 1000 $\mu$m
with a power exponent of $-3.5$.  Outside 80\,au we obtain the 
dust temperature and density on the spherical mesh by interpolating the 
gas density and temperature  from   the  hydrodynamic  simulations using
an  {\sph} interpolation    with     a    cubic    spline     smoothing    
kernel \citep{monaghan_1992} and assume a uniform gas-to-dust ratio of 100.  Due to the accretion of \sph\ particles on the central star, we extrapolate the density inwards of 80\,au using a radial power-law 
with an exponent chosen to give a smooth transition from the results of the hydrodynamic modelling.  We assume a vertical 
Gaussian density distribution, whose scaleheight is calculated from the temperature given by Eq.\,\ref{eq:T}.  
The dust and gas temperatures are assumed to be identical.  
Using  these temperatures and densities on a spherical mesh, we 
calculate  images at $\lambda=1.3$\,mm using the raytracer in \textsc{radmc-3d}.

Synthetic  observations  for  ALMA  are calculated  using  the  Common
Astronomy Software Application v4.5 (CASA, \citealp{mcmullin_2007}).
Visibilities  are  calculated with  the  {\tt  simobserve} task  while
imaging is  performed with the {\tt  clean} task.  Since our goal is  not to fit the observations exactly,  but merely to show
the   morphological   similarities   between  our   models   and   the
observations, we do not use exactly the same $(u,v)$ co-ordinates as the
observations to  calculate our  visibilities.  Instead,  we
choose an antenna configuration in CASA (alma.out13) which results
in a  very similar synthesised beam  to that of the  real data: 
0.27\arcsec$\times$0.25\arcsec\  with  PA  =  $86\degr$.   To  ensure  our  synthetic
observations are as close to the  real data as possible, we use a bandwidth of 6.8\,GHz, assume  a
precipitable water  vapour column  of 2.7\,mm,  and a  total integration
time of 725 seconds.

\smallskip

We then  apply an unsharp  masking filter to  the images in  a similar
manner to \cite{Elias227_Science}.  This involves convolving the
image with a  Gaussian kernel with a full width half maximum (FWHM) of 0.33\arcsec\  and subtracting a scaled
version  of  the  result  from  the  original  image.  This acts to remove large scale emission and boost the contrast of small scale structures.

\subsection{Observational constraints}
\label{sec:obs}

We describe various observational constraints that apply to the Elias 2-27 system, which we use when testing the three possible scenarios.

\begin{figure*}
\centering

\includegraphics[width=\textwidth]{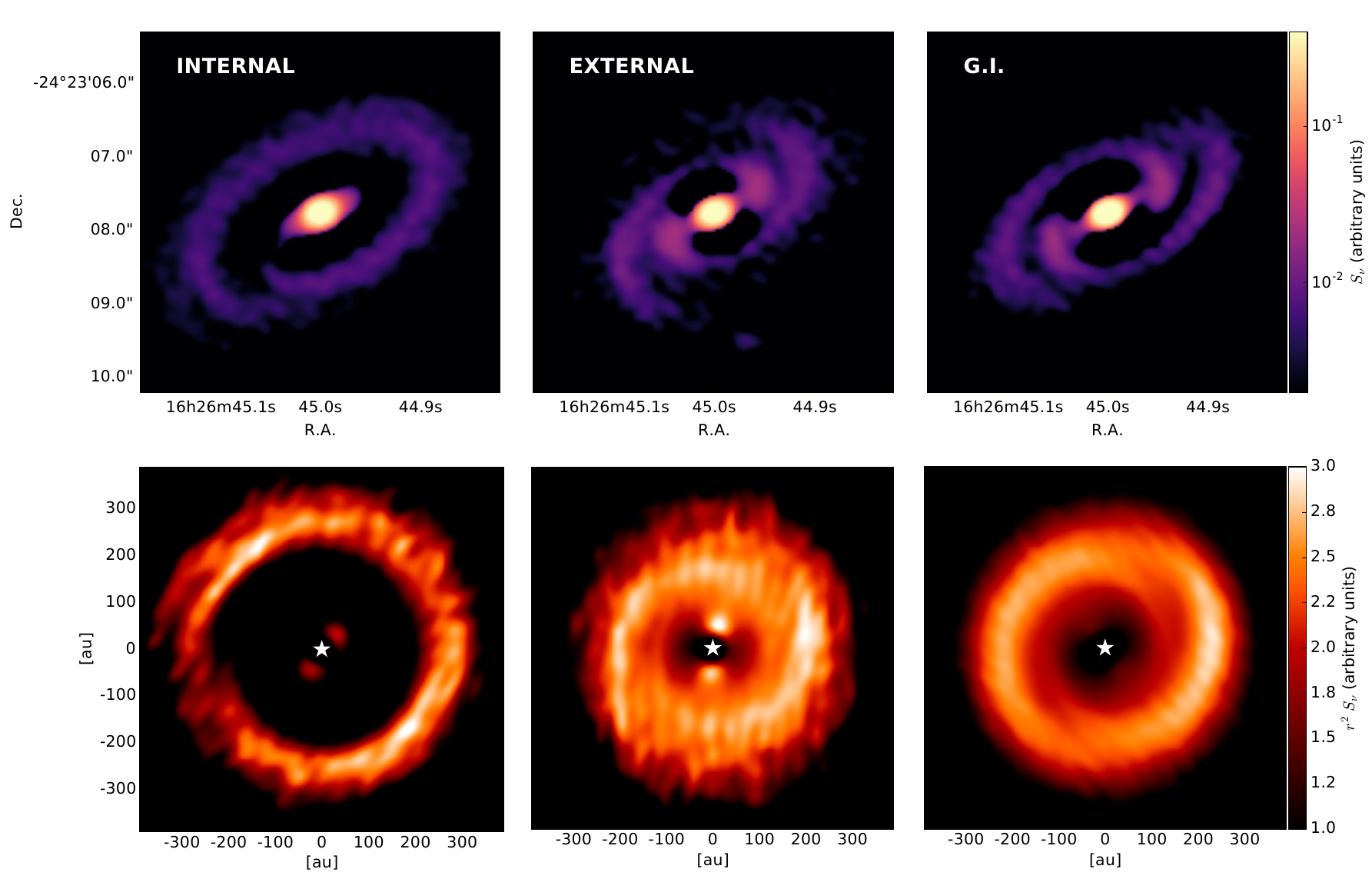}
\caption{Top: Unsharp   masked  images  of a disc with an internal companion (left), external companion (middle) and a gravitationally unstable disc (right).  The internal companion is not compatible with the observations.  The external companion simulations and the gravitationally unstable disc simulations show the closest match to the observation in Figure \ref{fig:ALMA1}.  Bottom: Deprojected mock ALMA images with an $r^{2}$ scaling of the simulations in the top panel (analogous to Figure~\ref{fig:ALMA1} right).  The intensity of the gravitationally unstable disc has been scaled down by a factor of 1.5 before the $r^{2}$ scaling is applied (see Section~\ref{sec:discussion}).  The simulations are run for 1.6, 1.2 and 3.75 orbits at 350au in the internal companion, external companion and self-gravitating disc simulations, respectively.}
\label{fig:results}
\end{figure*}

\subsubsection{Constraints on the total system mass}
\label{sec:msys}

The $^{12}$CO $J=2-1$ emission toward Elias 2-27 seems to be undergoing Keplerian rotation out to  $\sim 630$\,au, and according to kinematic modelling, the enclosed mass interior to the emission is $0.5 \pm 0.2 \Msolar$  \citep{Elias227_Science}.  However, absorption by the surrounding molecular cloud significantly obscures the red-shifted component of the emission, leading to some uncertainty in these derived masses.

\subsubsection{Constraints on the mass of a potential companion}
\label{sec:mcomp}

A volume-limited multiplicity survey of the $\rho$-Ophiuci molecular cloud was performed by \cite{ratzka_2005}. Elias\,2-27 was found to be a single star with an upper  limit on the K-band contrast of 2.5\,mag between  0.13--$6.4\arcsec$.  Additionally, the area around Elias 2-27 has been targeted by the UKIRT Infrared Deep Sky Survey (UKIDSS, e.g. \citealt{lawrence_ukidss_2007}).  The closest point source is located approximately $14\arcsec$ to the South ($16^{\rm h}26^{\rm m}44.95^{\rm s}$, $-24\degr23'21.88''$) with a K-band magnitude of 15.9\,mag, giving a K-band contrast of 7.5 mag with Elias 2-27.  The limiting magnitude of the survey in the K-band (17.8\,mag) suggests a maximum contrast with Elias 2-27 of 9.4\,mag for any undetected sources. 

\smallskip

Using the NextGen atmosphere models \citep{allard_1997,baraffe_1998,hauschildt_1999}, we convert these K-band contrasts to upper limits for the mass of any potential companion, $M_{\rm comp}$, for various orbital distances from Elias 2-27.  Assuming all objects lie at the same distance as Elias 2-27 (139\,pc), the UKIDSS data excludes any unseen companions $> 0.01 \Msolar$ beyond 420au (with the exception of the closest source mentioned above, which would translate to $M_{\rm comp} \approx 0.02\Msolar$ at 2000\,au).  Inside 420au a companion up to $0.08 \Msolar$ could be present based on \cite{ratzka_2005}.

\cite{Elias227_Science} present $^{12}{\rm CO}$, $\rm C^{18}O$ and $^{13}{\rm CO}$ observations which combined, show the presence of gas out to $\approx 630$~au.  Based on their channel maps and the PV diagram of $^{12}{\rm CO}$, there is no strong indication of a gap in the gas (though we note that the signal-to-noise is low).  In order to not open up a gap in the gas a planet must satisfy the viscosity and pressure conditions for gap-opening (Equations 68 and 69 of \citealp{Lin_Papaloizou_PPIII}; see also \citealp{Crida_gap}).  Between $300 - 420$~au (i.e. from the spirals to where the UKIDSS observations become relevant), the pressure condition is more stringent for $\alpha \lesssim 0.015$ providing a gap-opening mass of $\approx 13\MJup$ (though this mass limit is likely to be higher for migrating planets; \citealp{Malik_Meru_a}).  Beyond 420au the UKIDSS limit is more stringent than any realistic gap-opening mass.  Figure~\ref{fig:elias_diagram} shows a schematic diagram illustrating the observational and theoretical constraints.

%%%%%%%%%%%%%%%%%%%%%%%%%%%%%%%%%%%%%%%%%%%%%%%%%%%%%%%%%%%%%%%%%%%%%%%
\section{Results}
\label{sec:results}
%%%%%%%%%%%%%%%%%%%%%%%%%%%%%%%%%%%%%%%%%%%%%%%%%%%%%%%%%%%%%%%%%%%%%%%

We require the unsharp masked synthetic observations of the simulated discs to display morphology that is consistent with the observations.  This constitutes three main features --- \textit{i)}  two  large-scale  symmetric  spiral arms, \textit{ii)} two dark crescents interior to the spirals, and
\textit{iii)}  a  bright  inner  ellipse along the major axis of the disc (see Figure \ref{fig:ALMA1}, left).  The spiral arms are visible in the unsharped masking and deprojected images while the dark crescents and bright inner ellipse are only present in the unsharped masking image.  The discs presented in Figure~\ref{fig:results} employ a steep  outer disc  edge  rather  than an  exponentially  tapered  disc.

\subsection{Internal companion}
\label{sec:res_int}

Figure \ref{fig:results} (left) shows the simulated observation for one of our internal companion simulations.  The disc is a $0.08 \Msolar$ disc with $\Sigma \propto R^{-0.75}$ and $T \propto R^{-0.75}$, and includes a $0.01 \Msolar$ companion at 140\,au that is allowed to accrete from the disc and grow to approximately $0.03 \Msolar$.  The companion clears a large gap in the disc, forming the required central elliptical feature, but without the dark crescents or two armed spiral as in the original unsharped masking observations.  A lower companion mass does not generate the large-scale spirals, while higher mass companions remove large amounts of material from the disc.  We therefore suggest that the morphology in Elias 2-27 is unlikely to be due to an undetected companion internal to the spirals.

\subsection{External companion}

Figure \ref{fig:results} (centre) shows the simulated observations for one of our external companion simulations.  This disc is the same as that in Section~\ref{sec:res_int} but includes a $\approx 10 \MJup$ companion at $\approx 425$~au.  The simulated unsharped masking observation reproduces the large-scale spiral arms, the dark crescents and the bright inner ellipse, while the deprojected image shows two large-scale spirals analogous to Figure~\ref{fig:ALMA1} (right).  We also note that simulations with companions located much beyond the gas disc ($\gtrsim 700$~au) can only reproduce the observed morphology by violating the companion mass limits in Section \ref{sec:mcomp}.

\subsection{Gravitationally unstable disc}

Figure \ref{fig:results} (right) shows the simulated observation for a gravitationally unstable disc with $\Sigma \propto R^{-0.5}$ and $T \propto R^{-0.75}$.  The disc mass is 0.24 $\Msolar$, leading to a disc-to-star mass ratio of 0.49.  Even with this relatively massive disc, the combined system mass lies within the limits discussed in Section \ref{sec:msys}.  The simulated unsharp masking observation reproduces the large-scale spiral arms, the dark crescents and the bright inner ellipse, while the deprojected image shows two large-scale spirals analogous to Figure~\ref{fig:ALMA1} (right).

%%%%%%%%%%%%%%%%%%%%%%%%%%%%%%%%%%%%%%%%%%%%%%%%%%%%%%%%%%%%%%%%%%%%%
\section{Discussion}
\label{sec:discussion}
%%%%%%%%%%%%%%%%%%%%%%%%%%%%%%%%%%%%%%%%%%%%%%%%%%%%%%%%%%%%%%%%%%%%%

\begin{figure}
\centering
\includegraphics[width=1.0\columnwidth]{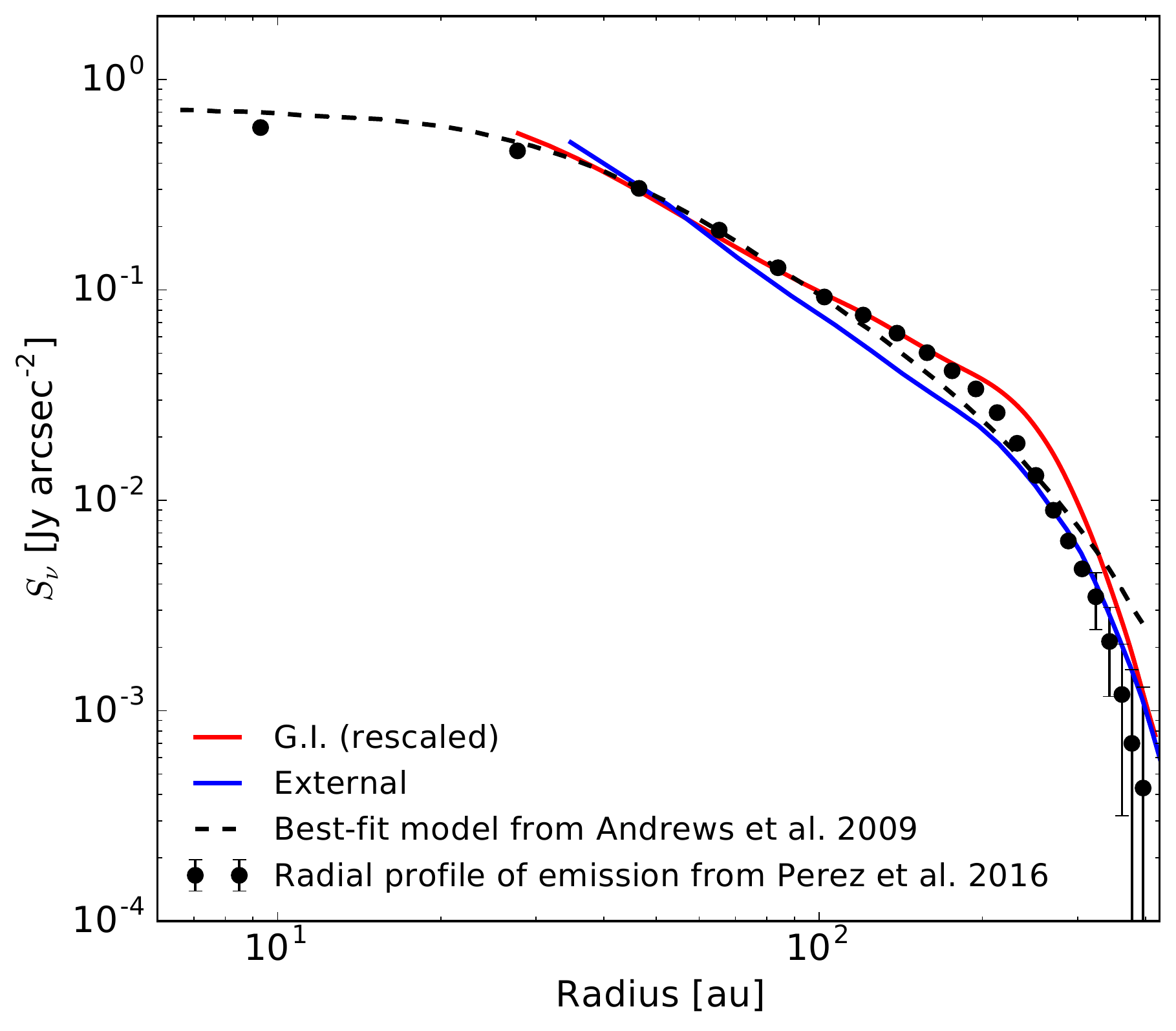}
\caption{Radial emission profiles for the successful gravitationally unstable disc simulation (red), which has been decreased by a factor of $\sim 1.5$ (see text) and the successful companion disc simulation (blue).  Also shown are data from \cite{Elias227_Science} (black points) and the best fit model from \citet{andrews_2009_oph} (black dash).  The steep decline beyond $\approx$~300\,au is key to matching the observed morphology in the simulated images.}
\vspace{2em}
\label{fig:intensity}
\end{figure}

The combination of our hydrodynamic modelling and simulated observations allows us to put strong constraints on the disc structure from which the sub-millimetre continuum emission originates.  Throughout all of our models, we find that a steeply declining surface mass density beyond $\approx$~300\,au is key to producing a close match to the masked image.  Figure~\ref{fig:intensity} shows the scaled radial intensity profile of our successful self-gravitating disc and the radial intensity profile of the successful companion simulations compared to the observational data from \cite{Elias227_Science} and the model of \citet{andrews_2009_oph}.  The steep decline beyond $\approx$~300\,au in our simulation manifests itself in the form of a steep emission profile at large radii that turns over at approximately the right radius, matching the observed profile well.  The modest scaling factor for the self-gravitating disc ($\sim 1.5$) is well within uncertainties associated with dust-to-gas ratios and/or grain opacities for circumstellar discs.

\smallskip

We note that the resulting Toomre profile dips marginally below 1 for a limited radial range in our successful self-gravitating disc.  However, this disc shows no sign of fragmentation even though we evolve it for many dynamical times for this radial range.  On the other hand a companion beyond the spirals ($\gtrsim 300$~au) is unlikely to have formed by core accretion.  This suggests that, should the companion hypothesis be correct, this would hint that the Elias 2-27 disc may have formed a fragment by gravitational instability in the past.

\smallskip

\begin{figure*}
\centering
\includegraphics[height=5.7cm]{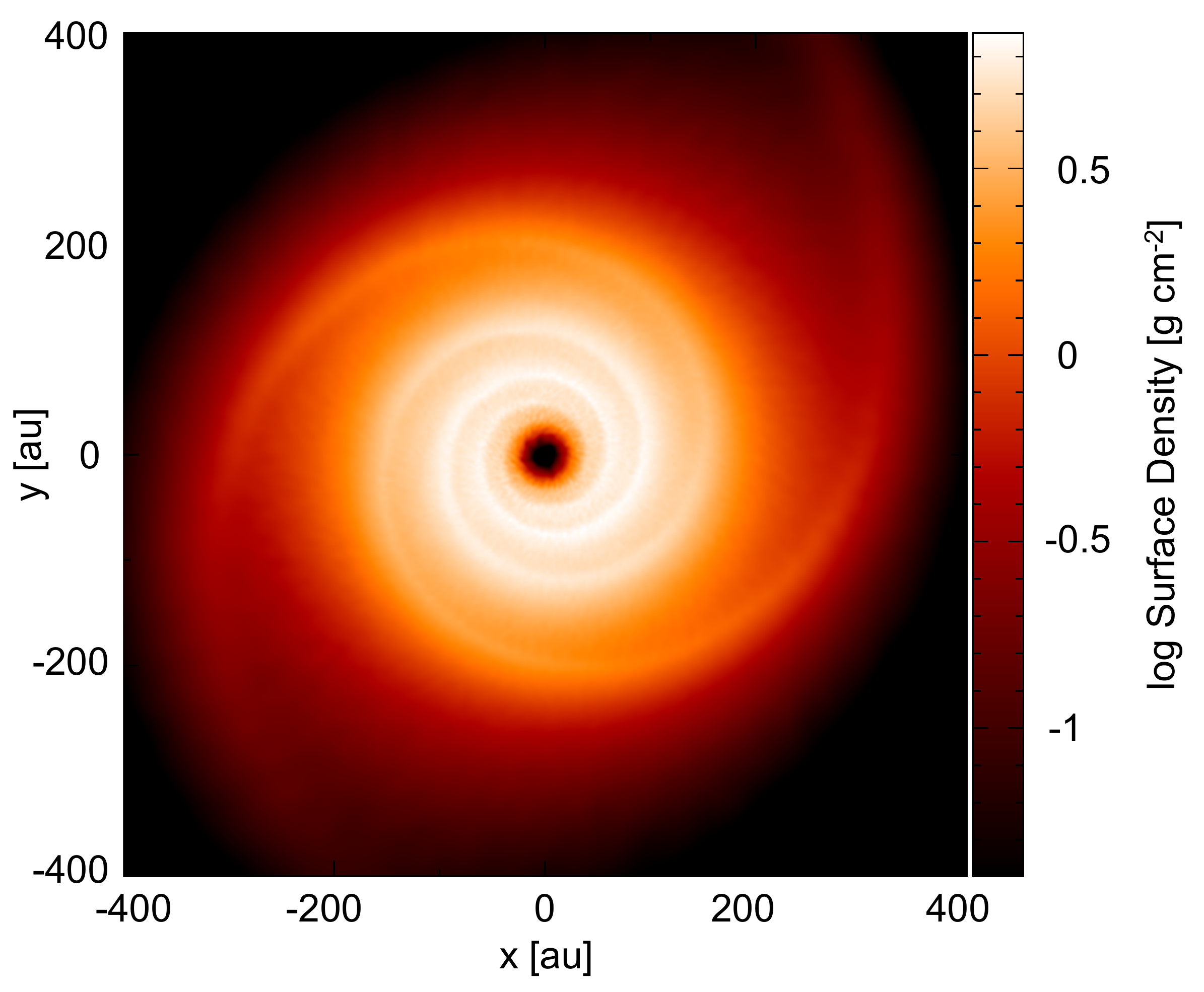}
\hspace{0.8cm}
\includegraphics[height=5.7cm,trim={0 0 0 0},clip]{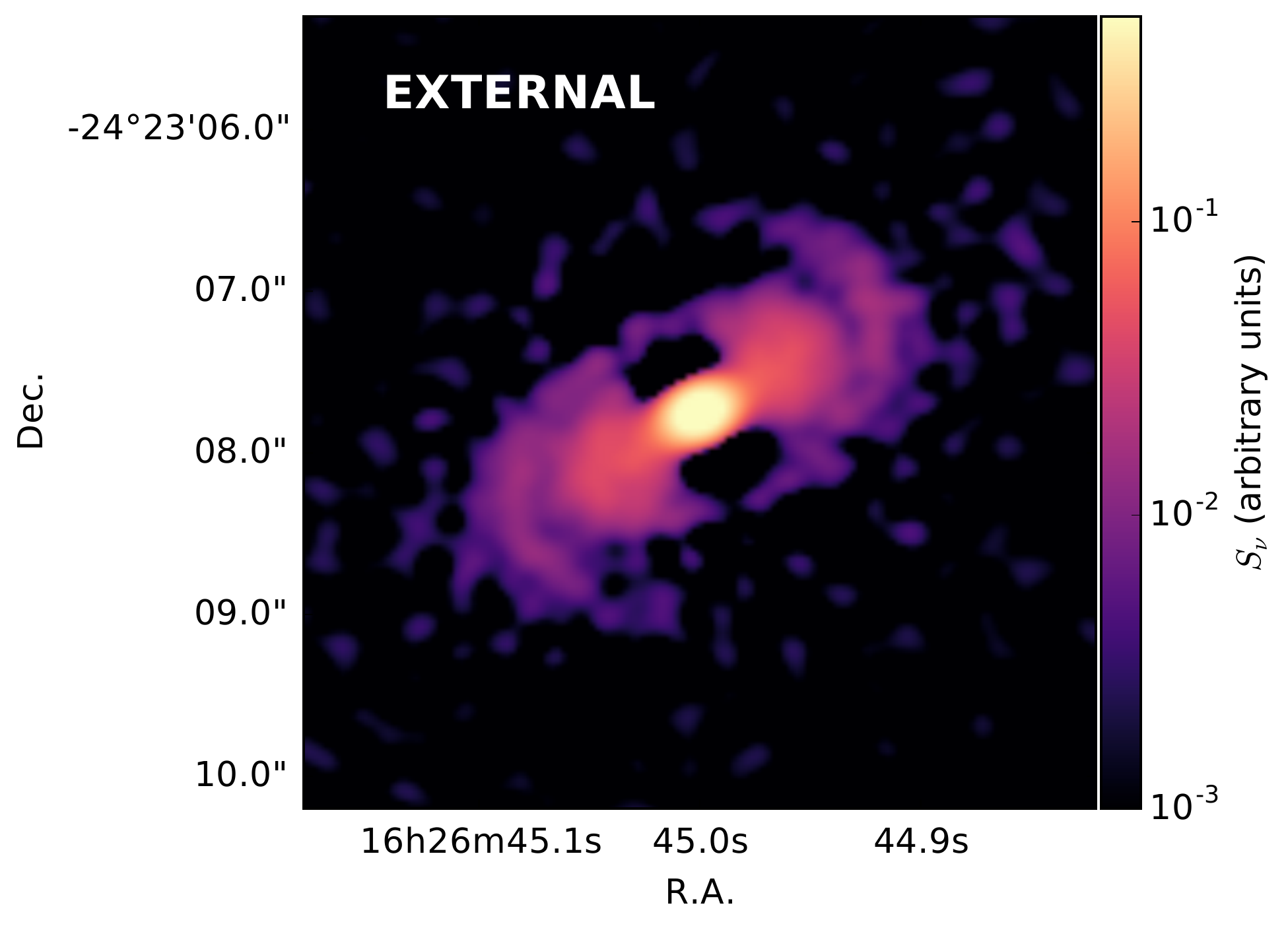} \\
\vspace{0.8cm}
\caption{Left: surface density of an external companion simulation (see Section \ref{sec:discussion} for details) which displays prominent two-armed spirals.  This simulation involved a $0.5 \Msolar$ companion on a circular orbit at $R = 1200$~au and was run for 1.3 orbits at 1200au.  Right: unsharp masked image of this simulation, which is unable to reproduce the morphology due to higher levels of emission in the inner disc.}
\label{fig:bad}
\end{figure*}

We also stress the importance of applying an unsharp mask to our simulated observations.  While such a mask is primarily applied to increase the contrast of the image, subtle changes are introduced to the resulting image that allow us to remove models from consideration.  Changes in the surface density profile affect the radial locations at which there is an excess or deficit with respect to a Gaussian mask. This is illustrated in Figure~\ref{fig:bad} which shows a disc modelled using the properties presented by \cite{Elias227_Science} (see Section~\ref{sec:hydro} for details) i.e. with an exponential surface mass density profile, which does not reproduce the observations well.  Therefore while a range of models replicate the spirals seen in the unprocessed ALMA image \citep[e.g.][]{Tomida_Elias227_GI}, comparison of masked images constrains the disc properties further.

\smallskip

However, special care must be taken when interpreting unsharp masked images.  This is because some structures may not be real, but artificially created by the masking.  For example, while Figure~\ref{fig:ALMA1} (left) superficially suggests that a {\it gap} exists in the disc, similar structures in the masked images of the external companion and self-gravitating disc in Figure~\ref{fig:results} are generated by the interaction of the mask with a \emph{smoothly declining} emissivity profile.  Masked images of observations therefore need to be interpreted via the type of forward modelling exercise undertaken here.  

\smallskip

A two-armed spiral structure, as observed around Elias 2-27, is consistent with high disc-to-star mass ratios \citep[e.g.][]{Lodato_Rice_massive_disc}.  Indeed \cite{Tomida_Elias227_GI}, who suggested that the Elias 2-27 disc is self-gravitating, obtained a high disc-to-star-mass ratio in their simulations.
The shocks  associated with spiral arms in self-gravitating discs have been shown to have an effect on the chemistry of the  disc material  \citep{ilee_2011,hincelin_2013,evans_2015} leading to the  prospect of detecting  the features  in line emission \citep[][]{douglas_2013}.  Future observations of the Elias 2-27 system, with sensitivities high enough to spatially resolve relevant molecular line transitions (e.g. CO, HCO$^{+}$, OCS and H$_2$CO) will be crucial in further evaluating the dynamics occurring within the star-disc system.

Finally, deep near infrared imaging should offer a first step towards deciding which scenario is at work in the system. If no companion is detected, gravitationally instability is likely. Otherwise further follow-up observations would be required to confirm any possible detected companion.

\smallskip

%%%%%%%%%%%%%%%%%%%%%%%%%%%%%%%%%%%%%%%%%%%%%%%%%%%%%%%%%%%%%%%%%%%%
\section{Conclusions}
\label{sec:conc}
%%%%%%%%%%%%%%%%%%%%%%%%%%%%%%%%%%%%%%%%%%%%%%%%%%%%%%%%%%%%%%%%%%%%

We present the results of a series of hydrodynamic and radiative transfer models to test three hypotheses regarding the origin of the disc morphology around Elias 2-27 --- a companion internal or external to the spirals, and a gravitationally unstable disc.  Our results show that a steep decline in surface mass density beyond $\approx 300$~au is required, and that the gravitational instability hypothesis or a $\lesssim 10 - 13\MJup$ companion between $\approx 300 - 700$~au can reproduce all components of the observed morphology.  Given this, we suggest that Elias 2-27 may be one of the first examples of an observed self-gravitating disc or a disc that has recently fragmented forming a $\lesssim 10 - 13 \MJup$ planet.

\acknowledgments We thank Mike Irwin for obtaining UKIDSS photometry, and Jim Pringle and the referee for useful comments.  We acknowledge support from the DISCSIM project, grant agreement 341137 under  ERC-2013-ADG.   FM acknowledges   support  from   The  Leverhulme   Trust.   This  paper  uses  the  following  ALMA  data: ADS/JAO.ALMA\# 2013.1.00498.S.  This work used the Darwin DiRAC HPC cluster at the University of Cambridge, and the Cambridge COSMOS SMP system funded by ST/J005673/1, ST/H008586/1 and ST/K00333X/1 grants.


\begin{thebibliography}{}
\expandafter\ifx\csname natexlab\endcsname\relax\def\natexlab#1{#1}\fi
\providecommand{\url}[1]{\href{#1}{#1}}

\bibitem[{{Allard} {et~al.}(1997){Allard}, {Hauschildt}, {Alexander}, \&
  {Starrfield}}]{allard_1997}
{Allard}, F., {Hauschildt}, P.~H., {Alexander}, D.~R., \& {Starrfield}, S.
  1997, \araa, 35, 137

\bibitem[{{ALMA Partnership} {et~al.}(2015){ALMA Partnership}, {Brogan},
  {P{\'e}rez}, {Hunter}, {Dent}, {Hales}, {Hills}, {Corder}, {Fomalont},
  {Vlahakis}, {Asaki}, {Barkats}, {Hirota}, {Hodge}, {Impellizzeri}, {Kneissl},
  {Liuzzo}, {Lucas}, {Marcelino}, {Matsushita}, {Nakanishi}, {Phillips},
  {Richards}, {Toledo}, {Aladro}, {Broguiere}, {Cortes}, {Cortes}, {Espada},
  {Galarza}, {Garcia-Appadoo}, {Guzman-Ramirez}, {Humphreys}, {Jung}, {Kameno},
  {Laing}, {Leon}, {Marconi}, {Mignano}, {Nikolic}, {Nyman}, {Radiszcz},
  {Remijan}, {Rod{\'o}n}, {Sawada}, {Takahashi}, {Tilanus}, {Vila Vilaro},
  {Watson}, {Wiklind}, {Akiyama}, {Chapillon}, {de Gregorio-Monsalvo}, {Di
  Francesco}, {Gueth}, {Kawamura}, {Lee}, {Nguyen Luong}, {Mangum}, {Pietu},
  {Sanhueza}, {Saigo}, {Takakuwa}, {Ubach}, {van Kempen}, {Wootten},
  {Castro-Carrizo}, {Francke}, {Gallardo}, {Garcia}, {Gonzalez}, {Hill},
  {Kaminski}, {Kurono}, {Liu}, {Lopez}, {Morales}, {Plarre}, {Schieven},
  {Testi}, {Videla}, {Villard}, {Andreani}, {Hibbard}, \&
  {Tatematsu}}]{ALMA_HLTau}
{ALMA Partnership}, {Brogan}, C.~L., {P{\'e}rez}, L.~M., {et~al.} 2015, \apjl,
  808, L3

\bibitem[{{Andrews} {et~al.}(2009){Andrews}, {Wilner}, {Hughes}, {Qi}, \&
  {Dullemond}}]{andrews_2009_oph}
{Andrews}, S.~M., {Wilner}, D.~J., {Hughes}, A.~M., {Qi}, C., \& {Dullemond},
  C.~P. 2009, \apj, 700, 1502

\bibitem[{{Andrews} {et~al.}(2016){Andrews}, {Wilner}, {Zhu}, {Birnstiel},
  {Carpenter}, {P{\'e}rez}, {Bai}, {{\"O}berg}, {Hughes}, {Isella}, \&
  {Ricci}}]{Andrews_TWHydra}
{Andrews}, S.~M., {Wilner}, D.~J., {Zhu}, Z., {et~al.} 2016, \apjl, 820, L40

\bibitem[{{Baraffe} {et~al.}(1998){Baraffe}, {Chabrier}, {Allard}, \&
  {Hauschildt}}]{baraffe_1998}
{Baraffe}, I., {Chabrier}, G., {Allard}, F., \& {Hauschildt}, P.~H. 1998, \aap,
  337, 403

\bibitem[{{Bate} {et~al.}(1995){Bate}, {Bonnell}, \&
  {Price}}]{Bate_Bonnell_Price_sink_ptcls}
{Bate}, M.~R., {Bonnell}, I.~A., \& {Price}, N.~M. 1995, \mnras, 277, 362

\bibitem[{{Benz}(1990)}]{Benz1990}
{Benz}, W. 1990, in Numerical Modelling of Nonlinear Stellar Pulsations
  Problems and Prospects,, ed. {J.~R.~Buchler} (Kluwer, Dordrecht), 269

\bibitem[{{Birnstiel} {et~al.}(2013){Birnstiel}, {Dullemond}, \&
  {Pinilla}}]{Birnstiel_dust_conc_time}
{Birnstiel}, T., {Dullemond}, C.~P., \& {Pinilla}, P. 2013, \aap, 550, L8

\bibitem[{{Booth} \& {Clarke}(2016)}]{Booth_Clarke_GIdust}
{Booth}, R.~A., \& {Clarke}, C.~J. 2016, \mnras, 458, 2676

\bibitem[{{Casassus} {et~al.}(2013){Casassus}, {van der Plas}, {M}, {Dent},
  {Fomalont}, {Hagelberg}, {Hales}, {Jord{\'a}n}, {Mawet}, {M{\'e}nard},
  {Wootten}, {Wilner}, {Hughes}, {Schreiber}, {Girard}, {Ercolano}, {Canovas},
  {Rom{\'a}n}, \& {Salinas}}]{casassus_2013}
{Casassus}, S., {van der Plas}, G., {M}, S.~P., {et~al.} 2013, \nat, 493, 191

\bibitem[{{Clarke} \& {Lodato}(2009)}]{Clarke_GI_planetesimals}
{Clarke}, C.~J., \& {Lodato}, G. 2009, \mnras, 398, L6

\bibitem[{{Crida} {et~al.}(2006){Crida}, {Morbidelli}, \& {Masset}}]{Crida_gap}
{Crida}, A., {Morbidelli}, A., \& {Masset}, F. 2006, \icarus, 181, 587

\bibitem[{{Douglas} {et~al.}(2013){Douglas}, {Caselli}, {Ilee}, {Boley},
  {Hartquist}, {Durisen}, \& {Rawlings}}]{douglas_2013}
{Douglas}, T.~A., {Caselli}, P., {Ilee}, J.~D., {et~al.} 2013, \mnras, 433,
  2064

\bibitem[{{Evans} {et~al.}(2015){Evans}, {Ilee}, {Boley}, {Caselli}, {Durisen},
  {Hartquist}, \& {Rawlings}}]{evans_2015}
{Evans}, M.~G., {Ilee}, J.~D., {Boley}, A.~C., {et~al.} 2015, \mnras, 453, 1147

\bibitem[{{Evans} {et~al.}(2009){Evans}, {Dunham}, {J{\o}rgensen}, {Enoch},
  {Mer{\'{\i}}n}, {van Dishoeck}, {Alcal{\'a}}, {Myers}, {Stapelfeldt},
  {Huard}, {Allen}, {Harvey}, {van Kempen}, {Blake}, {Koerner}, {Mundy},
  {Padgett}, \& {Sargent}}]{evans_2009_c2d}
{Evans}, II, N.~J., {Dunham}, M.~M., {J{\o}rgensen}, J.~K., {et~al.} 2009,
  \apjs, 181, 321

\bibitem[{{Hauschildt} {et~al.}(1999){Hauschildt}, {Allard}, \&
  {Baron}}]{hauschildt_1999}
{Hauschildt}, P.~H., {Allard}, F., \& {Baron}, E. 1999, \apj, 512, 377

\bibitem[{{Hincelin} {et~al.}(2013){Hincelin}, {Wakelam}, {Commer{\c c}on},
  {Hersant}, \& {Guilloteau}}]{hincelin_2013}
{Hincelin}, U., {Wakelam}, V., {Commer{\c c}on}, B., {Hersant}, F., \&
  {Guilloteau}, S. 2013, \apj, 775, 44

\bibitem[{{Ilee} {et~al.}(2011){Ilee}, {Boley}, {Caselli}, {Durisen},
  {Hartquist}, \& {Rawlings}}]{ilee_2011}
{Ilee}, J.~D., {Boley}, A.~C., {Caselli}, P., {et~al.} 2011, \mnras, 417, 2950

\bibitem[{{Isella} {et~al.}(2009){Isella}, {Carpenter}, \&
  {Sargent}}]{isella_2009_structure}
{Isella}, A., {Carpenter}, J.~M., \& {Sargent}, A.~I. 2009, \apj, 701, 260

\bibitem[{Isella {et~al.}(2016)Isella, Guidi, Testi, Liu, Li, Li, Weaver,
  Boehler, Carperter, De~Gregorio-Monsalvo, Manara, Natta, P\'erez, Ricci,
  Sargent, Tazzari, \& Turner}]{isella_2016}
Isella, A., Guidi, G., Testi, L., {et~al.} 2016, Phys. Rev. Lett., 117, 251101.
\newblock \url{http://link.aps.org/doi/10.1103/PhysRevLett.117.251101}

\bibitem[{{Lawrence} {et~al.}(2007){Lawrence}, {Warren}, {Almaini}, {Edge},
  {Hambly}, {Jameson}, {Lucas}, {Casali}, {Adamson}, {Dye}, {Emerson},
  {Foucaud}, {Hewett}, {Hirst}, {Hodgkin}, {Irwin}, {Lodieu}, {McMahon},
  {Simpson}, {Smail}, {Mortlock}, \& {Folger}}]{lawrence_ukidss_2007}
{Lawrence}, A., {Warren}, S.~J., {Almaini}, O., {et~al.} 2007, \mnras, 379,
  1599

\bibitem[{{Lin} \& {Papaloizou}(1993)}]{Lin_Papaloizou_PPIII}
{Lin}, D.~N.~C., \& {Papaloizou}, J.~C.~B. 1993, in Protostars and Planets III,
  ed. E.~H. {Levy} \& J.~I. {Lunine}, 749--835

\bibitem[{{Lodato} \& {Rice}(2005)}]{Lodato_Rice_massive_disc}
{Lodato}, G., \& {Rice}, W.~K.~M. 2005, \mnras, 358, 1489

\bibitem[{{Luhman} \& {Rieke}(1999)}]{rho_oph_spectraltype_age}
{Luhman}, K.~L., \& {Rieke}, G.~H. 1999, \apj, 525, 440

\bibitem[{{Malik} {et~al.}(2015){Malik}, {Meru}, {Mayer}, \&
  {Meyer}}]{Malik_Meru_a}
{Malik}, M., {Meru}, F., {Mayer}, L., \& {Meyer}, M. 2015, \apj, 802, 56

\bibitem[{{McMullin} {et~al.}(2007){McMullin}, {Waters}, {Schiebel}, {Young},
  \& {Golap}}]{mcmullin_2007}
{McMullin}, J.~P., {Waters}, B., {Schiebel}, D., {Young}, W., \& {Golap}, K.
  2007, in Astronomical Society of the Pacific Conference Series, Vol. 376,
  Astronomical Data Analysis Software and Systems XVI, ed. R.~A. {Shaw},
  F.~{Hill}, \& D.~J. {Bell}, 127

\bibitem[{{Meru}(2015)}]{Triggered}
{Meru}, F. 2015, \mnras, 454, 2529

\bibitem[{{Monaghan}(1992)}]{monaghan_1992}
{Monaghan}, J.~J. 1992, \araa, 30, 543

\bibitem[{{Morris} \& {Monaghan}(1997)}]{MorrisMonaghan1997}
{Morris}, J.~P., \& {Monaghan}, J.~J. 1997, J.\ Comp.\ Phys., 136, 41

\bibitem[{{Natta} {et~al.}(2006){Natta}, {Testi}, \&
  {Randich}}]{rho_oph_accretion}
{Natta}, A., {Testi}, L., \& {Randich}, S. 2006, \aap, 452, 245

\bibitem[{{Paardekooper} \& {Mellema}(2006)}]{Paardekooper2006_dust}
{Paardekooper}, S.-J., \& {Mellema}, G. 2006, \aap, 453, 1129

\bibitem[{{P{\'e}rez} {et~al.}(2016){P{\'e}rez}, {Carpenter}, {Andrews},
  {Ricci}, {Isella}, {Linz}, {Sargent}, {Wilner}, {Henning}, {Deller},
  {Chandler}, {Dullemond}, {Lazio}, {Menten}, {Corder}, {Storm}, {Testi},
  {Tazzari}, {Kwon}, {Calvet}, {Greaves}, {Harris}, \&
  {Mundy}}]{Elias227_Science}
{P{\'e}rez}, L.~M., {Carpenter}, J.~M., {Andrews}, S.~M., {et~al.} 2016,
  Science, 353, 1519

\bibitem[{{Price} \& {Bate}(2007)}]{Price_Bate_MHD_h}
{Price}, D.~J., \& {Bate}, M.~R. 2007, \mnras, 377, 77

\bibitem[{{Ratzka} {et~al.}(2005){Ratzka}, {K{\"o}hler}, \&
  {Leinert}}]{ratzka_2005}
{Ratzka}, T., {K{\"o}hler}, R., \& {Leinert}, C. 2005, \aap, 437, 611

\bibitem[{{Ricci} {et~al.}(2010){Ricci}, {Testi}, {Natta}, \&
  {Brooks}}]{ricci_2010_oph}
{Ricci}, L., {Testi}, L., {Natta}, A., \& {Brooks}, K.~J. 2010, \aap, 521, A66

\bibitem[{{Shi} \& {Chiang}(2014)}]{Shi_Chiang2014}
{Shi}, J.-M., \& {Chiang}, E. 2014, \apj, 789, 34

\bibitem[{{Tobin} {et~al.}(2016){Tobin}, {Kratter}, {Persson}, {Looney},
  {Dunham}, {Segura-Cox}, {Li}, {Chandler}, {Sadavoy}, {Harris}, {Melis}, \&
  {P{\'e}rez}}]{Tobin_Nature_GI_fragment}
{Tobin}, J.~J., {Kratter}, K.~M., {Persson}, M.~V., {et~al.} 2016, \nat, 538,
  483

\bibitem[{{Tomida} {et~al.}(2017){Tomida}, {Machida}, {Hosokawa}, {Sakurai}, \&
  {Lin}}]{Tomida_Elias227_GI}
{Tomida}, K., {Machida}, M.~N., {Hosokawa}, T., {Sakurai}, Y., \& {Lin}, C.~H.
  2017, \apjl, 835, L11

\bibitem[{{Weingartner} \& {Draine}(2001)}]{weingartner_2001}
{Weingartner}, J.~C., \& {Draine}, B.~T. 2001, \apj, 548, 296

\bibitem[{{Whitehouse} \& {Bate}(2006)}]{WH_Bate_science}
{Whitehouse}, S.~C., \& {Bate}, M.~R. 2006, \mnras, 367, 32

\bibitem[{{Whitehouse} {et~al.}(2005){Whitehouse}, {Bate}, \&
  {Monaghan}}]{WH_Bate_Monaghan2005}
{Whitehouse}, S.~C., {Bate}, M.~R., \& {Monaghan}, J.~J. 2005, \mnras, 364,
  1367

\bibitem[{{Zhu} {et~al.}(2012){Zhu}, {Nelson}, {Dong}, {Espaillat}, \&
  {Hartmann}}]{Zhu_dust_filtration}
{Zhu}, Z., {Nelson}, R.~P., {Dong}, R., {Espaillat}, C., \& {Hartmann}, L.
  2012, \apj, 755, 6

\end{thebibliography}
\end{document}